\documentclass[aps,pre,twocolumn,showpacs]{revtex4}
\usepackage{graphicx}
\usepackage{amssymb}
\usepackage{amsmath}
\usepackage{epsfig}

\begin{document}

\title{Long Time Evolution of Phase Oscillator Systems}

\author{Edward Ott and Thomas M. Antonsen}

\affiliation
{Institute for Research in Electronics and Applied Physics,
University of Maryland, College Park, MD 20742}


\begin{abstract}
It is shown, under weak conditions, that the dynamical evolution of an important class
of large systems of globally coupled, heterogeneous frequency, phase
oscillators is, in an appropriate physical sense, 
time-asymptotically attracted toward a reduced
manifold of system states. This manifold, which is invariant under the
system evolution, was previously known and used to facilitate the
discovery of attractors and bifurcations of such systems.
The result of this paper
establishes that attractors for the order parameter dynamics obtained by restriction
to this reduced manifold are, in fact, the only such attractors
of the full system. Thus {\it all} long time dynamical behavior
of the order parameters of these systems can be obtained 
by restriction to the reduced manifold.
\end{abstract}

\maketitle

{\bf
Systems consisting of many coupled phase oscillators have been used to model
diverse situations ranging 
from Josephson junction circuits, to circadian rhythms, to 
synchronization of cardiac pacemaker cells. In previous work 
by us \cite{Ott}, 
it was shown that a large class of such models possess
solutions on an invariant manifold {\it M}. It has since 
proved possible to simply obtain various
attractors of the dynamics on {\it M}. A remaining open question 
is that of whether such attractors for the dynamics on {\it M} are
also attractors for the dynamics of the full system, and, if so,
whether all of the attractors of the full system lie on {\it M}.
In this paper we prove, under very general conditions, that,
in an appropriate sense, the answer to these questions is yes. This
result establishes that restriction of consideration to the manifold
{\it M} can be used as an effective computational and analysis 
method for obtaining all the typical, long-time dynamical behavior of 
these systems.}

\section{Introduction}
Large systems of coupled phase oscillators with heterogeneous frequency
distributions are of general interest and are the essential modeling
tool in past analyses of a variety of interesting situations
in physics, chemistry, biology, etc. Perhaps the simplest and best
known such system is the Kuramoto model \cite{Kuramoto}, which treats
the synchronization of globally (all-to-all) coupled phase oscillators
for which the coupling between pairs of oscillators appears as the sine
of the phase difference between the oscillators. Examples where this basic
framework has been extended to more complex situations include Josephson
junction circuits \cite{Marvel}, pedestrian induced oscillation
of footbridges \cite{Abdulrehem,Eckhardt}, systems with time-dependent
coupling \cite{So}, 
driven systems describing circadian rhythm in mammals \cite{Antonsen,Childs},
the effect of time-delay in oscillator interactions \cite{Choi,Lee},
the effect of non-unimodal distribution of the natural frequencies 
of the phase oscillators \cite{Crawford,Martens}, ``communities'' of
phase oscillators interacting with multiple other phase oscillator
communities \cite{Barreto,Pikovsky}, the ``chimera'' model of certain
mammals that experience sleep with only one of their two brain 
hemispheres at a time \cite{Pikovsky,Abrams}, etc.

The large number of interesting applications of phase oscillator models
motivates the attempt to find general analysis tools applicable
to these models. In this vein, it has recently been shown that,
in the continuum limit (i.e., the number of oscillators approaches
infinity), such models possess solutions on a reduced 
manifold of system states \cite{Ott}. Furthermore, for the case
of a Lorentzian distribution of oscillator frequencies, the
dynamics on the reduced manifold is typically describable by a finite
number of ordinary differential equations. This finding has been utilized
to determine attractors and their bifurcations on the reduced manifold
for all the applications previously mentioned 
(see Refs.\cite{Marvel,Abdulrehem,So,Childs,Lee,Martens,Pikovsky,Abrams}). Two 
basic questions remain: 
(i) are attractors for the dynamics restricted
to the reduced manifold also attractors of the full system; and 
(ii) are there attractors of the full system that do not lie on the reduced
manifold?
Indications of results so far are mixed. On the one hand, numerical results
from Refs.\cite{Antonsen,Lee}, and especially \cite{Martens}, are 
consistent with the supposition that all attractors of the full system
lie on the reduced manifold. On the other hand, Refs.\cite{Pikovsky, Watanabe} find
long-time asymptotic behavior that is {\it not} on the reduced manifold.
The result of our paper is that, in an appropriate sense (that we specify
later in this paper), the reduced manifold is globally attracting provided 
that the spread $\Delta$ in the distribution of oscillator frequencies is 
nonzero. In particular, for $\Delta > 0$
all attractors of the full system lie on the reduced
manifold, and all attractors of the dynamics on the reduced manifold are 
attractors of the full system. This greatly facilitates the task of finding
the attractors of the full system, since they now can be sought using
the reduced system. The result also resolves the puzzle posed
by the previous results, since the finding by Refs.\cite{Pikovsky,Watanabe} of long 
time motion not on the reduced manifold was for the case of $\Delta = 0$, 
while the opposite indication from the numerical results 
of Refs.\cite{Antonsen,Lee,Martens} treated situations in which 
$\Delta > 0$.

\section{Formulation}
We begin by noting that the models in the class of problems in 
which we are interested all involve the determination of a distribution function 
$F(\theta,\omega,t)$, where $\theta$ is the phase of an oscillator,
and $\omega$ is the natural frequency an oscillator would have
in isolation from the outside world (e.g., from other oscillators);
$F d\theta d\omega$ is the fraction of oscillators at time $t$ whose phases
and natural frequencies lie in the range $[\theta,\theta + d \theta]$
and $[\omega,\omega + d \omega]$. Since the natural frequency $\omega$ of
an oscillator is assumed not to change with time, the marginal frequency
distribution, 
\begin{equation}
g(\omega) = \int_0^{2 \pi} F(\theta,\omega,t) d\theta,
\label{eq:marg_dist}
\end{equation}

\noindent
is time independent. The key quantity characterizing the macroscopic
behavior of the distribution function $F$ is the ``order parameter''
$r(t)$ originally introduced by Kuramoto \cite{Kuramoto} and 
defined by
\begin{equation}
r(t) = \int_{-\infty}^{\infty} \int_{0}^{2 \pi} 
        F(\theta,\omega,t) e^{-i \theta} d\theta d\omega.
\label{eq:ord_par}
\end{equation}

\noindent
Since the number of oscillators is conserved, $F$ obeys an oscillator
continuity equation,
\begin{equation}
\frac{\partial F}{\partial t} + \frac{\partial}{\partial \theta}
\left( v_{\theta} F \right) = 0,
\label{eq:osc_cont}
\end{equation}

\noindent
and, for all of the problems previously
mentioned (Refs.\cite{Marvel} to \cite{Abrams}),
$v_{\theta}(\theta,t)$ is expressible in the form \cite{Marvel},
\begin{equation}
v_{\theta} (\theta,\omega,t) = \omega + 
\frac{1}{2 i} \left[ H(t) e^{-i \theta} - H^*(t) e^{i \theta} \right].
\label{eq:v_the}
\end{equation}

\noindent
Equations (\ref{eq:osc_cont}) and (\ref{eq:v_the}) constitute an $\omega$-dependent
partial differential equation in the
two real variables ($\theta,t$) to be solved
subject to an initial condition $F(\theta,\omega,0)$. The problem is
ostensibly complicated by the fact that the time dependence of the quantity
$H$ may, in general (Refs.\cite{Kuramoto} to \cite{Ott}), depend on
$F$ through the complex order parameter $r(t)$ defined by (\ref{eq:ord_par}),
as well as through other, non-phase-oscillator variables obeying auxiliary 
dynamical equations, which themselves may \cite{Marvel,Abdulrehem} depend on $r(t)$, 
or (as in Ref. \cite{So}) through explicit external time dependence of system parameters.
Here are some
examples: for the classical Kuramoto \cite{Kuramoto} problem, 
$H = k r(t)$, where $k$ is the strength of the coupling between oscillators;
for the circadian rhythm problem \cite{Antonsen,Childs},
$H = k r(t) + \Lambda$, where $\Lambda$ is a constant reflecting the strength
of the diurnal drive of the day-night sunlight cycle, and 
$\Lambda$ might be given an explicit time dependence, $\Lambda = \Lambda(t)$,
to model variation between sunny and cloudy days; for the case of 
time delay in the response of oscillators to other oscillators in the
system \cite{Choi,Lee}, $H = k \int_0^{\infty} \rho(\tau) r(t-\tau) d\tau$,
where $\rho(\tau)$ is the distribution function 
\cite{Lee} of delays along the links 
between oscillators; in the cases treated 
in Refs.\cite{Barreto,Pikovsky} (communities
of oscillators), \cite{Martens} (nonunimodal frequency distribution
$g(\omega)$), and \cite{Pikovsky,Abrams} (the chimera model), there are
several distribution functions, i.e., $F$ and $H$ in (\ref{eq:osc_cont})
and (\ref{eq:v_the}) are replaced by $F_{\sigma}$ and $H_{\sigma}$
($\sigma = 1,2,\cdots,s$, where $s$ is the number of distributions) and 
each $H_{\sigma}$ is a function of all the order parameters 
$r_1,r_2,\cdots,r_s$; in the case of pedestrian induced oscillation of
footbridges \cite{Abdulrehem,Eckhardt}, $H=k \ddot{y}(t)$, where
$\ddot{y}(t)$ is the side-to-side acceleration of the bridge, which obeys
a damped oscillator equation driven by $r(t)$, where $r(t)$ represents
the effects of the pedestrians.

For a general problem of the type described above, as the system evolves,
$H(t)$ will change self consistently with $t$. We will show in what follows that,
{\it whatever} is the evolution of $H(t)$, in the long time limit,
solutions for the order parameter evolution $r(t)$ (or $r_{\sigma}(t)$)
obey the differential equation that applies for evolution on the reduced 
manifold of Ref.\cite{Ott}. Because the precise time dependence of 
$H(t)$ will not matter in the derivation of our result, it suffices 
to consider $H(t)$ as some general function of time without regard to
how this time dependence is determined.

Expanding the distribution $F$ as a Fourier series in $\theta$, we write
$F$ in the form,
\begin{equation}
F(\theta,\omega,t) = \frac{g(\omega)}{2 \pi} 
         \left[ 1 + F_+(\theta,\omega,t) + F_-(\theta,\omega,t) \right],
\label{eq:FourS}
\end{equation}

\begin{eqnarray}
F_+(\theta,\omega,t) &=& \sum_{n=1}^{\infty} F_n(\omega,t) e^{i n \theta},
\nonumber \\
F_-(\theta,\omega,t) &=& \sum_{n=1}^{\infty} F^*_n(\omega,t) e^{-i n \theta}.
\label{eq:FourSA}
\end{eqnarray}

\noindent
We note that $\int_0^{2 \pi} F_{\pm} d\theta = 0$ and that, assuming absolute
convergence of the Fourier series, the analytic continuation
of $F_+$ ($F_-$) into $Im(\theta)>0$ ($Im(\theta)<0$) has no singularities
and decays exponentially to zero as $Im(\theta) \rightarrow + \infty$
($Im(\theta) \rightarrow - \infty$). As will soon become evident, 
the decomposition of $F$ given by (\ref{eq:FourS}) is a key step. We note
that since $F_- = F_+^*$ for real $(\theta,\omega)$, it suffices to consider 
only $F_+$. Substituting (\ref{eq:FourS}) into 
Eqs.(\ref{eq:osc_cont},\ref{eq:v_the}) and projecting the result onto the 
function space spanned by the basis functions, 
$e^{i \theta},e^{2 i \theta},e^{3 i \theta}, \cdots$, we obtain
\begin{equation}
\frac{\partial F_+}{\partial t} + \frac{\partial}{\partial \theta}
\left\{ \left[ \omega + \frac{1}{2 i} \left( H e^{-i \theta} - H^* e^{i \theta} 
      \right) \right] F_+ \right\} 
= \frac{1}{2} H^* e^{i \theta}
\label{eq:osc_cont2}
\end{equation}

\noindent
As previously noted, our result will not depend on 
the precise time dependence of  $H(t)$.
Thus, whatever is the dependence of $H(t)$, we can formally regard it as given.  
Adopting this viewpoint, Eq.(\ref{eq:osc_cont2}) is linear in $F_+$ with
an inhomogeneous driving term on the right hand side (namely, 
$\frac{1}{2} H^*\exp(i \theta)$). As such, we can write $F_+$ as 
\begin{equation}
F_+ = \hat{F_+} + \hat{F_+}'
\label{eq:F_+}
\end{equation}

\noindent
where $\hat{F_+}$ is a homogeneous solution to (\ref{eq:osc_cont2}) and
$\hat{F_+}'$ is an inhomogeneous solution. An inhomogeneous solution is given
by taking the Fourier coefficients of $\hat{F_+}'$ 
to be given by  $\hat{F_n}'(\omega,t) = [\alpha(\omega,t)]^n$, as proposed in 
Ref.\cite{Ott}. When this ansatz is used in (\ref{eq:osc_cont2}), it is 
found that (\ref{eq:osc_cont2}) is indeed satisfied if $\alpha(\omega,t)$
satisfies 
\begin{equation}
\frac{\partial \alpha}{\partial t} + i \omega \alpha
+ \frac{1}{2} (H \alpha^2 - H^*) = 0,
\label{eq:alp_de}
\end{equation}

\noindent
which, for each value of $\omega$, is an ordinary differential equation
in time $t$. We note further that, as shown in \cite{Ott}, 
$|\alpha(\omega,t)| < 1$ so that the summation of the Fourier series
for $\hat{F_+}'$ converges and yields 
\begin{equation}
\hat{F_+}' = \frac{\alpha e^{i\theta}}{1-\alpha e^{i \theta}}.
\label{eq:F_+p}
\end{equation}

\noindent
This form of $\hat{F_+}'$ specifies the reduced manifold found in 
Ref.\cite{Ott}. Thus, at long time, $F_+$ would tend to $\hat{F_+}'$
(i.e., $F$ would tend to the reduced manifold) if 
$\lim_{t \rightarrow \infty} \hat{F_+} = 0$. However, a simple counter
examples shows that this cannot always be true. In particular, 
if $H=0$, Eq.(\ref{eq:osc_cont2}) has homogeneous solutions,
\begin{equation}
\hat{F_+} = \sum_n A_n e^{i n (\theta - \omega t)}
\label{eq:F_+p_H0}
\end{equation}

\noindent
for {\it any} set $\{ A_n \}$ for which this series converges. Since
the magnitude of each term of the summation in (\ref{eq:F_+p_H0})
 is time-independent ($\omega$ is real),
$\hat{F_+}$ does not go to zero at $t \rightarrow \infty$. On the other hand,
we note that, as $t$ increases, the individual terms, $e^{in(\theta-\omega t)}$,
oscillate more and more rapidly in $\omega$. Thus for any such term
\begin{displaymath}
I_n = \int_{-\infty}^{\infty} g(\omega) A_n e^{i n (\theta - \omega t)}
      d \omega
\end{displaymath}

\noindent
decays exponentially in time for sufficiently smooth $g(\omega)$. For
example, our subsequent considerations will be for the case of a 
Lorentzian frequency distribution,
\begin{equation}
g(\omega) = \frac{1}{\pi} \frac{\Delta}{\omega^2 + \Delta^2}
  = \frac{1}{2 \pi i} \left( \frac{1}{\omega-i\Delta} - \frac{1}{\omega+i\Delta}
                      \right),
\label{eq:Lorent}
\end{equation}

\noindent
for which 
\begin{equation}
I_n = A_n e^{i n \theta - n \Delta t},
\label{eq:In_Lor}
\end{equation}

\noindent
which decays exponentially to zero as $t \rightarrow \infty$ provided that
$\Delta > 0$ (the case $\Delta=0$ corresponds to $g(\omega)$ being a 
delta function in $\omega$). [We remark that the mean value of $\omega$ has
been taken to be zero in (\ref{eq:Lorent}), but that no generality is lost
by this, as a mean value can be restored by the change of variables, 
$\theta' = \theta - \Omega t$, $\omega' = \omega + \Omega$.] Thus, while
we cannot expect to show that $\hat{F_+} \rightarrow 0$ as $t \rightarrow \infty$,
Eq.(\ref{eq:In_Lor}) suggests that this may not be necessary to obtain order
parameter dynamics that tend to the order parameter dynamics 
that applies on the reduced manifold. In particular,
noting from (\ref{eq:ord_par}), (\ref{eq:FourS}) and (\ref{eq:FourSA}) that
\begin{equation}
r(t) = \int_{-\infty}^{\infty} \int_0^{2 \pi} F_+ g e^{-i \theta} 
       d\theta d\omega,
\label{eq:ordF_+}
\end{equation}

\noindent
what we require is that 
\begin{equation}
\lim_{t \rightarrow \infty} f_+(\theta,t) = 0,
\label{eq:cond_f_+}
\end{equation}

\noindent
where
\begin{equation}
f_+(\theta,t) = \int_{-\infty}^{\infty} \hat{F_+}(\theta,\omega,t) g(\omega)
d\omega.
\label{eq:f_+}
\end{equation}

\noindent
In what follows we will demonstrate that Eq.(\ref{eq:cond_f_+}) indeed 
holds under very general conditions.

\section{Demonstration of the main result}
To show that (\ref{eq:cond_f_+}) applies, we now assume that
the analytic continuation of $\hat{F_+}(\theta,\omega,t)$ into
$Im(\omega)<0$ has no singularities in $Im(\omega)<0$ and approaches zero
as $Im(\omega) \rightarrow - \infty$. To show that this last assumption is a 
consistent one, let $|\omega|$ be very large, $|\omega| \gg H$. Then the
homogeneous version of Eq.(\ref{eq:osc_cont2}) for $\hat{F_+}$ is 
approximately
\begin{displaymath}
\frac{\partial \hat{F_+}}{\partial t} + \omega 
\frac{\partial \hat{F_+}}{\partial \theta} \approx 0,
\end{displaymath}

\noindent
which has solutions for its Fourier $\theta$-components
$\hat{F_n} \sim \exp[in(\theta - \omega t)]$ which go to zero as 
$Im(\omega) \rightarrow - \infty$. (It was to achieve this that we have
introduced the decomposition of $F$ given by Eqs.(\ref{eq:FourS}) and 
(\ref{eq:FourSA}).) We will further discuss this analyticity assumption at the end 
of this paper. 

We now specialize to the case of Lorentzian $g(\omega)$, Eq.(\ref{eq:Lorent}).
We multiply the homogeneous version of Eq.(\ref{eq:osc_cont2}) by 
$g(\omega)d\omega$, integrate the result from $\omega=-R$ to $\omega=+R$,
analytically continue into the complex $\omega$-plane, close the integration
path with a semicircle of radius $R$ in the lower half $\omega$-plane, and
let $R \rightarrow \infty$. Using our assumption that $\hat{F_+}(\theta,\omega,t)$
is analytic in the lower half $\omega$-plane and decays to zero as 
$Im(\omega) \rightarrow -\infty$, the integral along the large semicircle
approaches zero as $R \rightarrow \infty$, and the integrals from 
$\omega = -\infty$ to $\omega = +\infty$ along the real $\omega$-axis may thus
be evaluated as the residue of the enclosed pole of $g(\omega)$ at 
$\omega = - i \Delta$ (see Eq.(\ref{eq:Lorent})). This yields
\begin{equation}
\frac{\partial f_+(\theta,t)}{\partial t} +
\frac{\partial}{\partial \theta}  
\left[ v(\theta,t) f_+(\theta,t) \right] = 0,
\label{eq:osc_cont3}
\end{equation}

\begin{equation}
v(\theta,t) = -i \left[ \Delta + \frac{1}{2} \left( e^{-i\theta} H(t)
                        - e^{i\theta} H^*(t) \right) \right],
\label{eq:v_theta}
\end{equation}

\noindent
where $f_+(\theta,t) = \hat{F_+}(\theta,-i\Delta,t)$.

We now introduce a conformal transformation of the upper half complex 
$\theta$-plane into the unit disc, $z = e^{i \theta}$. Equations (\ref{eq:osc_cont3})
and (\ref{eq:v_theta}) then become
\begin{equation}
\frac{\partial \tilde{f_+}(z,t)}{\partial t} +
\frac{\partial}{\partial z}  
\left[ \tilde{v}(z,t) \tilde{f_+}(z,t) \right] = 0,
\label{eq:osc_cont4}
\end{equation}

\noindent
where
\begin{eqnarray}
\tilde{v}(z,t) &=& \Delta z + \frac{1}{2} \left( H(t) - z^2 H^*(t) \right),
\label{eq:v_z} \\
\tilde{f_+}(z,t) &=& f_+(\theta,t) / e^{i \theta}. 
\label{eq:ttf_+}
\end{eqnarray}

\noindent
Noting that (\ref{eq:osc_cont4}) can be written as
\begin{equation}
\frac{d \tilde{f_+}(z,t)}{dt} + \tilde{f_+}(z,t) 
    \frac{\partial \tilde{v}(z,t)}{\partial z} = 0,
\label{eq:osc_cont5}
\end{equation}

\noindent
where $d/dt = \partial / \partial t + \tilde{v} \partial / \partial z$,
we can integrate (\ref{eq:osc_cont5}) along the characteristics of this
equation to obtain
\begin{equation}
\tilde{f_+}(z,t) = \tilde{f_+}(Z(z,0),0) \exp 
  \left[ -\eta(z,t)
  \right] ,
\label{eq:tf_+}
\end{equation}
\noindent
where
\begin{displaymath}
\eta(z,t) =  \int_0^t \left( \frac{\partial \tilde{v}(z',t')}{\partial z'}
                   \right)_{z'=Z(z,t')} dt',
\end{displaymath}

\noindent
and the characteristics are given by the orbit equation,
\begin{equation}
\frac{d Z(z,t')}{dt'} = \tilde{v}(Z(z,t'),t'),
\label{eq:ZZ}
\end{equation}

\noindent
with the final condition $Z(z,t)=z$. Thus $Z(z,t')$ for $t'<t$ represents
the location $Z$ of the orbit $\tilde{v}(z',t')$ that winds up
at point $z$ at time $t$. It is useful to rewrite (\ref{eq:ZZ}) by
introducing 
\begin{displaymath}
Z = \rho e^{i \phi}, \hspace{2mm} H = h e^{i \beta}
\end{displaymath}

\noindent
with $h, \beta, \rho$ and $\phi$ real. The real and imaginary parts of 
(\ref{eq:ZZ}) then give 
\begin{eqnarray}
\frac{d\rho}{dt'} &=& \tilde{v}_\rho = \rho \Delta + \frac{h}{2}
   (1-\rho^2) \cos(\phi - \beta), 
\label{eq:drho} \\
\rho \frac{d\phi}{dt'} &=& \tilde{v}_\phi = - \frac{h}{2}
    (1+\rho^2) \sin(\phi - \beta).
\label{eq:dphi}
\end{eqnarray}

\noindent
We note from (\ref{eq:drho}) that when $\rho=1$, we have $d\rho/dt'=\Delta > 0$.
Thus for final conditions on $\rho=1$, the orbits backward in time move into
$\rho < 1$.
Thus $|Z(z,t')|<1$ for $|z| \leq 1$
and $t'<t$ (i.e., $Z(z,t')$ is in the unit disc). We wish to show that 
$\tilde{f_+}(z,t) \rightarrow 0$ as $t \rightarrow +\infty$. From 
(\ref{eq:tf_+}) we see that this will be the case if 
\begin{displaymath}
\lim_{t \rightarrow \infty} Re [ \eta(z,t) ] = + \infty.
\end{displaymath}

\noindent
In order to show this, we first note that
the real part of $\partial \tilde{v}(z',t')/\partial z'$ is simply 
one half the divergence of the two dimensional flow 
$\tilde{\bf v} = \tilde{v}_\rho (\rho,\phi) {\bf \rho_0}
+ \tilde{v}_\phi(\rho,\phi) {\bf \phi_0}$ (where
$\tilde{v}_\rho$ and $\tilde{v}_\phi$ are given by (\ref{eq:drho})
and (\ref{eq:dphi}), and 
${\bf \rho_0}$ and ${\bf \phi_0}$ are unit vectors in the $\rho$
and $\phi$ directions), i.e.,
\begin{equation}
Re \left(  \frac{\partial \tilde{v}(z',t')}{\partial z'} \right)
\Big|_{z'=Z(z,t')}  =
\frac{1}{2} \nabla \cdot {\bf\tilde{v}}.
\label{eq:tmp1}
\end{equation}

\noindent
Equation(\ref{eq:tmp1}) is most easily demonstrated in rectangular co-ordinates:
$z'=x+iy$, $\tilde{v}(z',t') = \tilde{v}_x(x,y,t') + i \tilde{v}_y(x,y,t')$, where
$\tilde{v}_x, \tilde{v}_y, x$ and $y$ are real. 
Then (\ref{eq:tmp1}) immediately follows by setting 
${\bf \tilde{v}} = \tilde{v}_x {\bf x_0} + \tilde{v}_y {\bf y_0}$, and
using the Cauchy-Riemann condition, $\partial \tilde{v}_x / \partial x 
= \partial \tilde{v}_y /
\partial y$, in the expression for the divergence in rectangular coordinates,
$\nabla \cdot {\bf \tilde{v}} = \partial \tilde{v}_x/\partial x 
+ \partial \tilde{v}_y/\partial y$.
Now evaluating $\nabla \cdot {\bf \tilde{v}}$ in polar coordinates
$(\rho,\phi)$, we have 
\begin{equation}
\nabla \cdot {\bf \tilde{v}} = 
\frac{1}{\rho} \frac{\partial}{\partial \rho}
(\rho \tilde{v}_\rho) + \frac{1}{\rho} \frac{\partial \tilde{v}_\phi}{\partial \phi}
= 2[\Delta - h \rho \cos(\phi-\beta)].
\label{eq:Div_v}
\end{equation}

\noindent
Solving (\ref{eq:Div_v}) for $h \cos(\phi-\beta)$ in terms of 
$\nabla \cdot {\bf \tilde{v}}$ and inserting the result in Eq.(\ref{eq:drho})
for $d\rho/dt'$, we obtain after some rearrangement 

\begin{align}
Re \left(  \frac{\partial \tilde{v}(z',t')}{\partial z'} \right)
\Big|_{z'=Z(z,t')}
= &\Delta \frac{1+\rho^2(z,t')}{1-\rho^2(z,t')} \nonumber \\
&+ \frac{d}{dt'} \ln[1-\rho^2(z,t')].
\label{eq:tmp2}
\end{align}

\noindent
Inserting (\ref{eq:tmp2}) into the integral for $\eta(z,t)$ and choosing
a fixed reference time $T$ satisfying $0<T<t$, we have
\begin{align}
Re[\eta(z,t)] = &\int_{t-T}^t Re \left( \frac{\partial \tilde{v}(z',t')}{\partial z'}
                 \right)_{z'=Z(z,t')} dt' \nonumber \\
      &+ \ln \left[ \frac{1-\rho^2(z,t-T)}{1-\rho^2(z,0)} \right]  \nonumber \\
      &+ \Delta \int_0^{t-T} \frac{1+\rho^2(z,t')}{1-\rho^2(z,t')} dt'.
\label{eq:eta_eval}
\end{align}

\noindent
We are interested in final ($t'=t$) conditions on the unit circle,
$Z(z,t)=z=e^{i \theta}$ for $\theta$ real, and their continuation into the
unit disc $|z| \leq 1$, corresponding to $\rho \leq 1$ at the final
time $t'=t$.
For $\rho$ sufficiently near one, Eq.(\ref{eq:drho}) shows that 
$d\rho/dt' \cong \Delta$. Thus by the continuity of the right hand
side of (\ref{eq:drho}), there is an annulus in 
the $Z$-plane, $1 \geq \rho \geq \rho_0$, in which 
$d\rho/dt' > 0$, implying that as $t'$ is reduced from 
$t$ (i.e., $t-t'$ is increased),
$\rho$ moves uniformly from $\rho=1$ at time $t$ to smaller 
values. Thus any final point in the annulus eventually enters the disc
$\rho<\rho_0<1$ and never leaves it. We can therefore choose
the time $T$ such that for all final conditions $|Z(z,t)|=|z|\leq 1$,
we have 
\begin{equation}
\rho(z,t') < \rho(z,t-T) < \rho_0 < 1
\label{eq:rho_eval}
\end{equation}

\noindent
where the first inequality applies for $0 \leq t'<t-T$. 
We consider $T$ to be held fixed, and we ask how $\eta(z,t)$ behaves as 
$t \rightarrow +\infty$. By (\ref{eq:Div_v}) the integrand in the
first of the three terms of (\ref{eq:eta_eval}) is bounded, and, since
the integration range in $t'$ (namely, $T$) for this term is
fixed, we conclude that the first term is bounded. By (\ref{eq:rho_eval})
the second term in (\ref{eq:eta_eval}) is also bounded. Again by 
(\ref{eq:rho_eval}) the integrand of the third term of 
(\ref{eq:eta_eval}) is positive and greater than one. Thus the third
term exceeds $(t-T)\Delta$. Hence 
$Re[\eta(z,t)] \rightarrow +\infty$ for
$t \rightarrow +\infty$, if $\Delta > 0$,
thus demonstrating that the exponential factor in (\ref{eq:tf_+})
goes to zero for large time \cite{note3}. 
We therefore conclude from 
(\ref{eq:eta_eval}) and (\ref{eq:tf_+}) that 
Eq.(\ref{eq:cond_f_+}) is satisfied if $\Delta > 0$,
which is the desired result.

In particular, at large $t$ the order parameter will approach the 
quantity $\int_{-\infty}^{\infty} \alpha(\omega,t) g(\omega) d\omega$,
where $\alpha(\omega,t)$ evolves by Eq.(\ref{eq:alp_de}). This implies
that $r(t)$ will satisfy the differential equation, 
\begin{equation}
\frac{dr(t)}{dt} + \Delta r(t) + \frac{1}{2} \left[
H(t) r^2(t) - H^*(t) \right] = 0,
\label{eq:dr}
\end{equation}

\noindent
which follows from multiplying (\ref{eq:alp_de}) by $g(\omega)d\omega$,
integrating from $\omega=-\infty$ to $\omega=+\infty$ and, as done
previously, using the residue method to evaluate the integrals. Hence,
for $\Delta > 0$, the long time dynamics of the order parameter $r(t)$ is
governed by the ordinary differential equation 
(Eq.(\ref{eq:dr}) and Ref.\cite{Ott}) 
that describes its dynamics for distribution functions
$F$ on the reduced manifold. This is our main result.

\section{Discussion}
The principal conditions for the applicability of our result are that the initial
condition is such that, when $F_+(\theta,\omega,t)$ is continued into the
complex $\omega$-plane, it is analytic in $Im(\omega)<0$ and decays to zero 
as $Im(\omega) \rightarrow -\infty$. As discussed in Ref.\cite{Ott}, if 
these conditions are satisfied initially, then they are also satisfied
for all $t>0$. What happens if the condition at $Im(\omega) \rightarrow -\infty$
is {\it not} satisfied
initially? Here, a simple example \cite{Ott2} may be instructive. We again
consider the case $H=0$. Say the initial condition on $F_+$ has a 
component $\exp(i n \theta + i \gamma \omega)$ with $\gamma$ real and
positive. This initial condition violates our assumption of decay to 
zero as $Im(\omega) \rightarrow -\infty$. However, use of this initial
condition in Eqs.(\ref{eq:osc_cont},\ref{eq:v_the}) with $H=0$
yields the solution $\exp(i n \theta + i(\gamma-t)\omega)$, and, for
large enough time, $t>\gamma$, the result satisfies the required condition 
that it approaches zero as 
$Im(\omega) \rightarrow -\infty$ \cite{note2}. Thus,
even if our desired condition at $Im(\omega) \rightarrow -\infty$ 
is not satisfied initially, in many cases, the result 
that the long time dynamics of $r(t)$ is described by Eq.(\ref{eq:dr}) may 
still apply.

We now connect our result with the concept of an {\it inertial manifold}. An
inertial manifold $M$ with respect to a distance metric $\mu$ 
satisfies the condition that, for any initial
condition in the state space, the subsequent system evolution is such that
the distance between the evolved orbit and the manifold $M$ 
{\it as measured by the metric} $\mu$ approaches zero as $t \rightarrow +\infty$.
What we have shown in this paper is that, in the space of distribution functions
$F(\theta,\omega,t)$, our reduced manifold (Eq.(\ref{eq:F_+p})) is inertial
with respect to the proper distance metric $\mu$. In particular,
this is so if we take the distance between two distribution functions
$F_1(\theta,\omega,t)$ and $F_2(\theta,\omega,t)$ to be defined by
\begin{equation}
\mu(F_1,F_2) = \left\{ \int_0^{2\pi} \left[ \int_{-\infty}^{\infty} (F_1-F_2)
    g(\omega) d\omega \right]^2 d\theta \right\}^{1/2}.
\label{eq:dist_met}
\end{equation}

\noindent
(For $F_1$ not on the reduced manifold $M$, the distance from $F_1$
to $M$ is $\mu(F_1,F_2)$ minimized over all $F_2$ on $M$.) Note that, by
this choice of distance metric, the problem associated with the example 
of Eq.(\ref{eq:F_+p_H0}) is avoided.

Finally, we remark that, while our result is for the special case of a
Lorentzian distribution of oscillator frequencies (Eq.(\ref{eq:Lorent})),
we believe that this restriction does not greatly limit the usefulness
of the resulting formulations for discovering typical system behavior. 
Indeed, past numerical experiments \cite{Antonsen,Martens} comparing
results obtained using Lorentzian $g(\omega)$ and using 
Gaussian $g(\omega)$ were
found to yield qualitatively identical bifurcation structures.

We are grateful to B.R. Hunt for an extremely useful comment. We also thank
W.S. Lee and S.H. Strogatz for comments on a preliminary draft of this
paper. This work was supported by the 
ONR(N00014-07-0734) and by NSF (PHY 0456249).



\begin{thebibliography}{1}
\bibitem{note3}
Equation (\ref{eq:ZZ}), together with the final condition $Z(z,t'=t)=z$
can be viewed as generating a conformal mapping from the complex
$z$-plane to the complex $Z$-plane. Since Eq.(\ref{eq:ZZ}) with 
$\tilde{v}(Z,t')$ given by (\ref{eq:v_z}) is a Riccati equation, this
mapping is a M\"{o}bius transformation, $Z=A(z-B)/(z-C)$,
where the coefficients $A, B$ and $C$ depend on $t$. [See,
for example, E. Hille, {\it Ordinary Differential Equations in the
Complex Domain}, page 105, Eq.(4.1.6) (Courier Dover, 1997)]. Thus the unit 
disc in $z$ (i.e., $|z| \leq 1$) is mapped into a disc in $Z$, and since, by
(\ref{eq:drho}), $d\rho/dt' = \Delta > 0$ at $\rho=1$ 
for $t' \leq t$, this disc is contained
within the region $|Z| \leq 1$. Our result, 
$Re[\eta(z,t)] \rightarrow +\infty$
as $t \rightarrow +\infty$, implies that
the radius of this $Z$-disc shrinks to zero as $t \rightarrow +\infty$.
\bibitem{note2}
It is interesting to note that $\int g(\omega) 
\exp[i n \theta + i(\gamma - t) \omega] d \omega = 
\exp(in\theta-|\gamma-t|\Delta)$ for Lorentzian $g(\omega)$, and that this
term {\it increases} exponentially with $t$ for $t < \gamma$, but
then {\it decreases} exponentially with $t$ for $t > \gamma$. This
general type of behavior is what is responsible for the ``echo''
phenomenon in Ref.\cite{Ott2}. We also note that the result that $f_+$ 
tends to zero even though $F_+$  need not is similar to behavior of 
the distribution function for linear, Landau damped, 
waves in collisionless plasmas. E.g., see Ref.\cite{Ott2} and
S.H.Strogatz, R.E.Mirollo and P.C.Matthews, Phys.Rev.Lett. {\bf 68}, 2730 (1992).
\end{thebibliography}
\end{document}